# Investigation of the time evolution of entanglement and trace distance in an atom-cavity system described with random walk and non-random walk states

M. Mohammadi[1], S. Jami[2]

[1,2]Department of physics, Mashhad Branch, Islamic Azad University, Mashhad, Iran

**Abstract**

An atom-cavity system consists of an atom or group of atoms inside a cavity. When an atom in cavity is stimulated by a laser pump, it is affected by the atom-field interaction shows the qusi- random walk behavior. This can change the entanglement and trace distance of system. In this work, the change entanglement and trace distance of system in two different cases -namely- the random walk and non-random walk is considered. The descriptive system is a two-level atom in the electrodynamics cavity based on the Jayne's-Cummings model, which is stimulated by two longitudinal and transverse laser pumps. The results show that the consideration of the random walk case for the atom changes in the amount of entanglement can be seen as the phenomenon of sudden death and birth of entanglement. It results in its rate of changes to be increased. In contrast to the non-random walk case, this rate is increased and decreased more quickly compared to the random walk case. The maximum amount of entanglement in each case is the same, but the minimum in random walk case is less than that of non-random walk case.

## 1. Introduction

One of the strangest topics in quantum mechanics is entanglement, which is the most fundamental concept that distinguishes classical physics from quantum physics [1-2]. The entanglement phenomena include many potential applications in communications and quantum computing, namely: quantum teleportation [3- 4], quantum-safe communications and quantum computing [5-6]. However, historically, until 1990, it was considered as the physical curiosity and strange feature without practical application. Ekert proposed the first effective application related to quantum entanglement since 1991 [7]. In the following years, Bennett and Wiesner[8-9] showed that two entangled parts can communicate between two classical-bits via sending only one qubit, for instance, a quantum system in two-dimensional Hilbert's space. The entangled are both conceptually and practically significant. One of the simplest schemes that can be made into quantum entanglement is the use of two-level atom systems. In the 1990s, the entangled containing two atoms were experimentally studied using ultra cold trapped ions [10-11] and quantum electrodynamics in a cavity (QED) [12]. It is possible to create quantum entanglement by stimulating the one or several atoms in an atom-cavity system using a laser field [13-14]. In this case, one can see the thermal and cooling effects in an atom-cavity system [15-19]. In this study, we have investigated the time evolution of entanglement corresponding to an atomic cavity system by considering the atomic oscillations.

In fact, in this research two laser pumps and along with the kinetic energy of the atom are considered. In this system, the electro-optical field is applied to the atom and cavity, imitates the atom to behave random-walk that can be attributed as a quantum dynamic due to the considering system [20-21]. Whether the atom is taken to be static or dynamic within the cavity, kinetic energy part can alter the entanglement. In the following, we obtain the entanglement in a fully quantum system considering two states, namely, random walk (RW) and non-random walk (NRW), respectively. Moreover, the results will be compared with each other.

In the second section, the model is introduced. In the third section the calculation of the entanglement related to the atom-cavity system is provided with the consideration of random walk and without random walk. In the fourth section, the results obtained by two states are compared to each other. In the fifth section, the calculation of the trace distance for the atom-cavity system is discussed and in the last section, the results are investigated.

## 2. Model description

The system under consideration contains an atom inside the cavity, which is affected by two laser pumps. The Hamiltonian for such a system includes non-interacting term; say that $H_0$, a term associated with the atomic interaction, $H_{a-f}$ and the last term is related to the pumps, $H_p$. We consider the desired system as a single two-level atom, although, in reality, the systems are multi-level atoms. However, once the atom is subjected to a field, the probability of transition to a level with a frequency close to that of the field is much higher than other levels. In this case, the other levels are ignored. Both the atoms and the field are taken as quantum identities. Hence, we consider the interaction of the field and the atom in a fully Jayne's-Cummings quantum model. This model is used to express the interaction between a field and an atom [22-24]. Experimental setups implement cavity and Rydberg atoms in a strong coupling regimen in the Jayne's- Cummings model [25-29]. The mathematical description of the Hamiltonian is in the following:

$$H = H_0 + H_{a-f} + H_P \tag{1}$$

$$H_0 = \frac{p^2}{2m} + \frac{\hbar \omega_a}{2} \hat{\sigma}^z + \hbar \omega_c \hat{a}^\dagger \hat{a} \tag{2}$$

$$H_{a-f} = i\hbar\, g\, f(x)\, (\hat{\sigma}^+ \hat{a} - \hat{a}^\dagger \hat{\sigma}^-) \tag{3}$$

$$H_p = i\hbar\, \eta_L\, (\hat{a}^\dagger e^{-i\omega_L t} - \hat{a}\, e^{i\omega_L t}) + i\hbar\, \eta_T\, (\hat{\sigma}^+ e^{-i\omega_T t} - \hat{\sigma}^- e^{i\omega_T t}) \tag{4}$$

Where p is the momentum operator of the atom, $\omega_a$ is its frequency, $\omega_c$ is the frequency corresponding a field in a cavity, $g(x) = g\, f(x)$ is the function for the atom-field coupling, the expression $f(x)=cos(kx)$ introduces the atomic displacement, $\eta_L$ and $\eta_T$ are the longitudinal and transverse laser pump, field amplitude with corresponding frequency $\omega_L$ and $\omega_T$, respectively. K is the wave vector, and $\hat{a}^\dagger$ ($\hat{a}$) are the creation (annihilation) operators, $\hat{\sigma}^z$ is the Pauli operator, and $\hat{\sigma}^-$ ($\hat{\sigma}^+$) are the raising (lowering) ladder operators.

The Lindblad equation can be evolved by the time-dependent Schrodinger equation in the following relation:

$$i\hbar \frac{d\rho}{dt} = [\rho, H_{eff}] \tag{5}$$

$$H_{eff} = H_s - i\hbar \sum \Gamma_i\, \alpha^\dagger{}_i \alpha_i \tag{6}$$

Where $\Gamma_i$ the decaying is rate, and $H_s$ is the Hamiltonian for the system in question. The $H_{eff}$ can be expressed by the following equation:

$$H_{eff} = \frac{p^2}{2m} + \frac{\omega_a}{2}\hat{\sigma}^z + \omega_c\hat{a}^\dagger\hat{a} + i\hbar\, g\, f(x)\,(\hat{\sigma}^+\hat{a} - \hat{a}^\dagger\hat{\sigma}^-) + i\hbar\, \eta_L(\hat{a}^\dagger e^{-i\omega_L t} - \hat{a}\, e^{i\omega_L t}) + i\hbar\, \eta_T(\hat{\sigma}^+ e^{-i\omega_T t} - \hat{\sigma}^- e^{i\omega_T t}) - i\hbar\, \Gamma_p \frac{p^2}{2m} - i\hbar\, \Gamma_c\, \hat{a}^\dagger\hat{a} - i\hbar\, \Gamma_a\, \hat{\sigma}^z \quad (7)$$

The solution for the Schrodinger equation is:

$$\psi_{total} = p(t)\,(c_1|e,n\rangle + c_2|g,n+1\rangle) \quad (8)$$

By rewriting the Schrodinger equation and act its parts on wave function separately, we have:

$$\dot{C}_1 = -i\Omega_c n C_1 - i\Omega_a C_1 + g\langle p|f(x)|p\rangle C_2\sqrt{n+1} - i\frac{\langle p^2\rangle}{2m}C_1 - \dot{p}C_1 \quad (9)$$

$$\dot{C}_2 = -i\Omega_c(n+1)C_2 + i\Omega_a C_2 - g\langle p|f(x)|p\rangle C_1\sqrt{n+1} - i\frac{\langle p^2\rangle}{2m}C_2 - \dot{p}C_2 \quad (10)$$

Where the expressions are: $\Omega_c = \omega_c - i\Gamma_c$, $\Omega_a = \frac{\omega_a}{2} - i\Gamma_a$, "n" is the quantum numbers of the photon and $C_1$ and $C_2$ are complex numbers. On the other hand, by using above relations the position and momentum derivatives are given by:

$$\dot{x} = 2\omega_r p \quad (11)$$

$$\dot{p} = -2g f'(x)\, Im\{\alpha^*\beta\} \quad (12)$$

Respectively ($\omega_r = \hbar^2 k^2/2m$ is the recoil frequency).

In the following, the time evolution of entanglement considering with RW comparing to NRW states will be discussed.

3. **Time evolution of quantum entanglement**

   **(A): The random walk state**

   In this state, the wave function for a given state is expressed as follows:

$$\psi_{total} = C_{11}|e,n\rangle + C_{12}|e,n+1\rangle + C_{21}|g,n\rangle + C_{22}|g,n+1\rangle \quad (13)$$

The concurrence relation provides the following equation for the coefficients:
$$C = 2\,|C_{11}C_{22} - C_{12}C_{21}| \quad (14)$$

The pure state for the system in question is written by:

$$\psi_{total} = p(t)(C_1|e,n\rangle + C_2|g,n+1\rangle) \quad (15)$$

In the previous section, the differential equation and the given coefficients $C_1$ and $C_2$ were already introduced.

Now we examine the entanglement over time, for two different initial states. An initial state is a system that is separate, and the other mode is with maximum entanglement or Bell state. Fig.1 shows the results of time evolution associated with these different initial states.

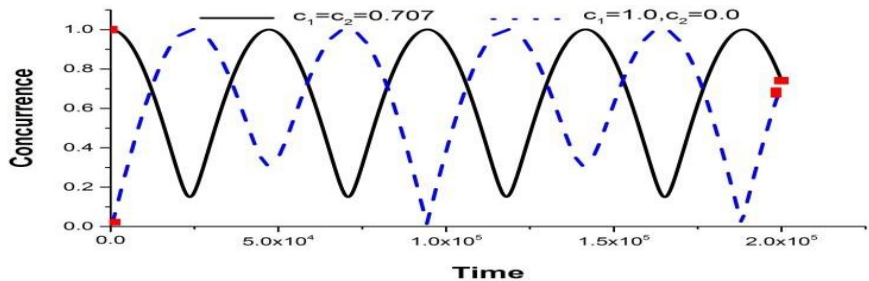

**Fig.1.** Comparison of entanglement with the concurrence criterion in two different initial states (considering the RW for the atom)

From Fig.1 it can be seen that the initial state with the maximum entanglement decreases over time, reaching a minimum value but not zero. It also continues with the same periodic state. But when the initial state is set to be without entanglement, the amount of entanglement starts from zero and increases over the time, and reaches maximum value and consequently, the amount of entanglement decreases and goes to zero over a period of time resulting the phenomenon of sudden death occurs.

**(B): The non-random walk state**

In this section, we want to consider that if the atomic momentum is not considered, following which the kinetic energy term is eliminated and how the entanglement of the system changes. By removing the kinetic term from eq. (2) for the wave function we obtain:

$$\psi_{total} = C_1|e,n\rangle + C_2|g,n+1\rangle \tag{16}$$

In this case, the effective Hamiltonian is expressed as follows:

$$H_{eff} = \frac{\hbar\omega_a}{2}\hat{\sigma}^z + \hbar\omega_c \hat{a}^\dagger \hat{a} + i\hbar\, g\, f(x)(\hat{\sigma}^+\hat{a} - \hat{a}^\dagger\hat{\sigma}^-) + i\hbar\,\eta_L(\hat{a}^\dagger e^{-i\omega_L t} - \hat{a}\, e^{i\omega_L t}) + i\hbar\,\eta_T(\hat{\sigma}^+ e^{-i\omega_T t} - \hat{\sigma}^- e^{i\omega_T t}) - i\hbar\,\Gamma_c \hat{a}^\dagger \hat{a} - i\hbar\,\Gamma_a \hat{\sigma}^z \tag{17}$$

We obtain the coefficients $C_1$ and $C_2$ in principle with the procedure used before. Then with respect to the concurrence relation for the two initial states, namely, separable state and Bell state, the system in question shows a behavior depicted in Fig.2.

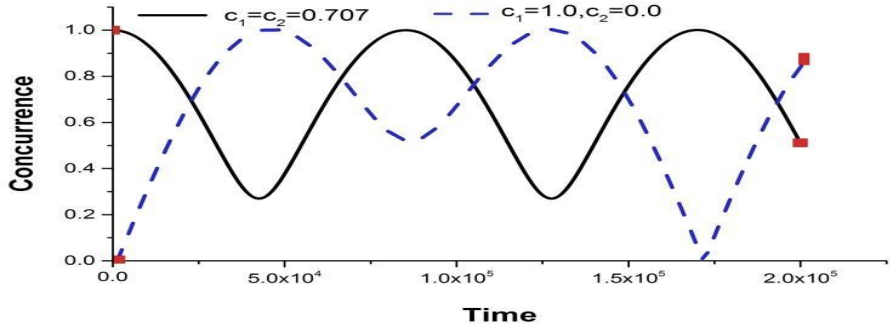

**Fig.2.** Comparison of entanglement with the concurrence criterion in two different initial states (considering the NRW atom).

Figure.2 shows that while the system is in maximum entanglement, the amount of its entanglement decreases over time and reaches a minimum value but does not go to zero. Also as one can observe, this state will continue oscillations over the time. However, as the system begins with a zero concurrence value, the amount of entanglement is increased over time, reaching a maximum value, and then decreased again. Here, the important point is that the amount of entanglement reaches a minimum of zero.

**4. Comparison of time evolution of entanglement with the concurrence criterion considering RW and NRW states for the atom**

Here the importance considering RW, and NRW states for the atom over the system in question is compared. In Section2, we discussed each case independently. If the entanglement in the atom-cavity system is set to be zero initially and one intends to find out the time evolution of entanglement with the concurrence criterion considering RW and NRW states for the atom, one can observe the behavior of the system shown in Fig.3.

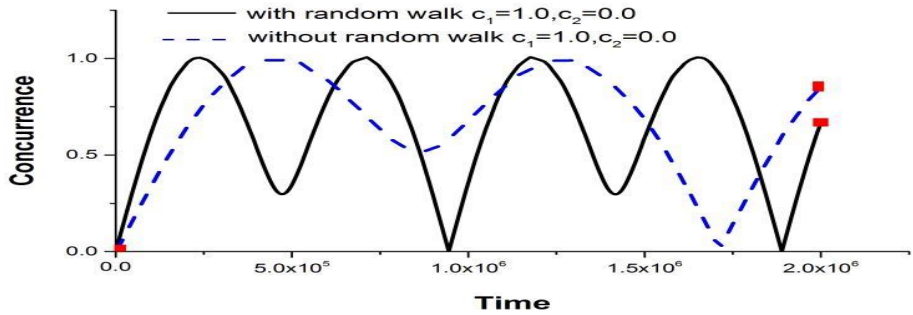

**Fig.3.** Comparison of entanglement with the concurrence criterion in two cases; (NRW and RW cases are shown with blue dash and black solid line, respectively).

Figure.3 indicates that in both cases, the amount of entanglement increases and reaches a maximum of unit over the time. In the RW case, the amount of entanglement is increased and the system reaches the maximum in the time interval. However, in the NRW case, the rate of increase in entanglement is slower, and the system reaches the maximum in a greater time. Over time, also it can be seen that the amount of entanglement is decreased. This decreasement occurs faster for the RW case compared to the NRW.

In NRW case, the rate of reduction in the amount of entanglement is lower and, after reducing to a minimum value, again gains the maximum value of entanglement and then reaches zero. This happens over a longer period of time. There are some time intervals in which the amount of entanglement is minimum, but for the case of NRW is maximum and vice versa. Increasing and decreasing the amount of entanglement for the case of RW is faster than that of NRW case. This can be referred to more interaction in the atom-cavity system. Now if the system is initially at its maximum entanglement and we are asked to find out the time evolution of the entanglement, the results are depicted in Fig.4.

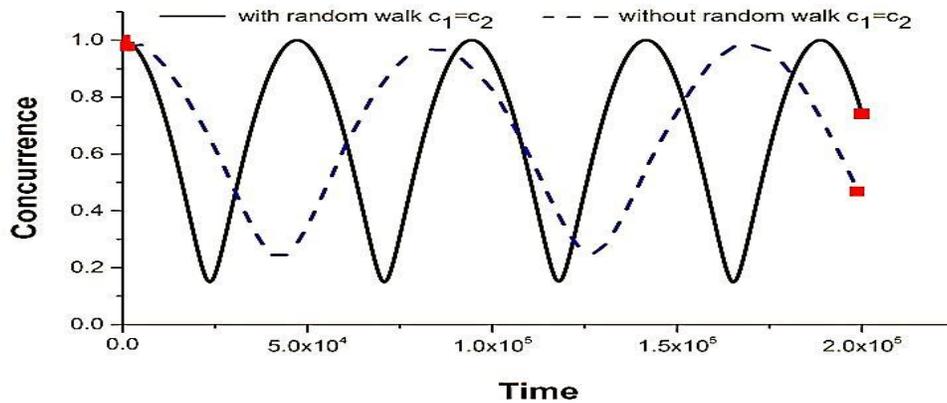

**Fig.4.** Comparison of entanglement with the concurrence criterion in both RW and NRW cases with primary condition of $C_1 = C_2 = 1/\sqrt{2}$.

Figure.4 shows the variation in entanglement for the atom-cavity system under condition that the system initially was in maximum entanglement. This condition is the same for both RW and NRW cases. In each case, the amount of entanglement is initially reduced. In the RW case, this rate of decrease is faster and happens in lesser time, then reaches a minimum value. While in the NRW, the rate of decrease is slower compared to the RW, i.e., reducing the rate of entanglement is slower, and the minimum value for this case is higher than that of RW case. Over time, the amount of entanglement is increased, and reaches a maximum amount of unit. As shown in Fig.4one can observe an oscillating behavior, but the amount of entanglement does not reach zero.

## 5. Calculation of trace distance of system

Given that the relationship between the density matrix and the trace distance according to quantum mechanics standard books is as follows [1-2]:

$$D(\rho,\sigma) = \frac{1}{2} \|\rho - \sigma\| = \frac{1}{2} [\sqrt{(\rho-\sigma)^\dagger(\rho-\sigma)}\,] \tag{18}$$

$$\rho_{atom-field} = |\varphi_{atom-field}\rangle\langle\varphi_{atom-field}| \tag{19}$$

Where $D(\rho,\sigma)$ and $\rho_{a-f}$ are trace distance and density matrix, respectively. On the other hand, the wave function for the system in question can be expressed as:

$$\varphi_{a-f} = C_1|e,n\rangle + C_2|g,n+1\rangle \tag{20}$$

By this expression, density matrix can be written as:

$$\rho_{a-f} = (C_1|e,n\rangle + C_2|g,n+1\rangle)(\langle e,n|C_1^* + \langle g,n+1|C_2^*) = \begin{pmatrix} C_1 C_1^* & C_1 C_2^* \\ C_1^* C_2 & C_2 C_2^* \end{pmatrix} \tag{21}$$

In this step, one intends to obtain the reduced density matrix. Therefore, the photon states along with field are traced over the field, so we have:

$$\rho_{atom} = \sum_m \langle m|\rho_{a-f}|m\rangle = (C_1 C_1^* |e,n\rangle\langle e,n|) + (C_2 C_2^*|g,n+1\rangle\langle g,n+1|)$$

$$= \begin{pmatrix} |C_1|^2 & 0 \\ 0 & |C_2|^2 \end{pmatrix} \tag{22}$$

If density matrix is divided into matrices $\sigma$ and $\rho$ (since the system in question is a pure) showing the system time in t=0 and a certain t, respectively, we can write as follow:

$$\sigma = \sigma(t=0) = \begin{pmatrix} |C_1(t=0)|^2 & 0 \\ 0 & |C_2(t=0)|^2 \end{pmatrix} \tag{23}$$

$$\rho = \rho(t) = \begin{pmatrix} |C_1(t)|^2 & 0 \\ 0 & |C_2(t)|^2 \end{pmatrix} \tag{24}$$

Noting the expressions above, the calculated trace distance for both RW and NRW cases is shown in Fig.5.

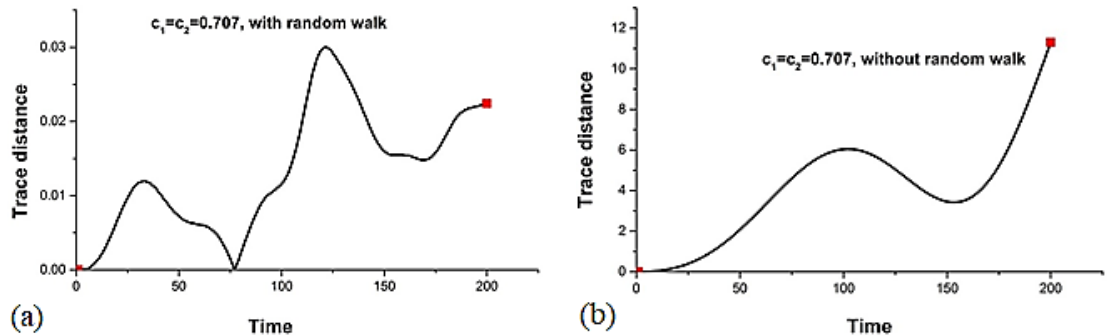

**Fig.5.** representation of calculated trace distance over time for (a) the random walk and (b) non-random walk cases with primary condition of $c_1 = c_2 = 1/\sqrt{2}$.

## 6. Conclusion

In this research, it was tried to investigate the quantum entanglement related to an atom-cavity system with consideration of atomic motion. The results show that the consideration of RW for the atomic motion changes the amount of entanglement and can be seen as the phenomenon of sudden death and appearance of entanglement which results in its rate of changes to be increased. In contrast to the NRW case, this rate is increased and decreased more quickly compared to the RW case. The maximum amount of entanglement in each case is the same, but the minimum is RW is less than that of NRW case. In the following, the trace distance is also obtained indicating the similarity between initial and final state. The importance of the trace distance can be highlighted by two facts: first, it shows that the atom imitates the random walk behavior and secondly, is indicative of the gate or the memory of the atom-cavity system. The calculation of trace distance determined that the atom does not remain in its original state and in some cases, it changes quickly and others it does slower.